\documentclass[prl,twocolumn,amsmath,amssymb,aps]{revtex4-2}

\usepackage{comment}
\usepackage{xcolor}
\usepackage{graphicx}
\usepackage{dcolumn}
\usepackage{bm}
\usepackage{hyperref}
\usepackage{kantlipsum}
\hypersetup{pdfborder=0 0 0,colorlinks=true,citecolor=blue,linkcolor=blue}

\begin{document}

\title{
Effects of intra-layer correlations on electron-hole double-layer superfluidity}
\author{Filippo Pascucci$^{1,2}$, Sara Conti$^3$, Andrea Perali$^{1}$, Jacques Tempere$^2$, and David Neilson$^3$}
\affiliation{$^1$Supernano Laboratory, School of Pharmacy, University of Camerino, 62032 Camerino (MC), Italy}
\affiliation{$^2$TQC, University of Antwerp, Universiteitsplein 1, 2610 Antwerp, Belgium}
\affiliation{$^3$CMT, Dept.\ of Physics, University of Antwerp, Groenenborgerlaan 171, 2020 Antwerp, Belgium}

\date{\today}

\begin{abstract}
We investigate the correlations acting within the layers in a superfluid system of electron-hole spatially separated layers. In this system of quasi-dipoles, the dominant correlations are Hartree--Fock. We find in the BEC regime of the superfluid where screening is negligible, that the effect of the correlations on superfluid properties is also negligible.  However, in the BCS-BEC crossover regime, where the screening plays a crucial role, we find that the superfluid gap is significantly weakened because the correlations significantly boost the number of low-energy particle-hole excitations participating in the screening process.
Finally, the intralayer correlations are found in this system to suppress a predicted phenomenon in which the average pair size passes through a minimum as the crossover regime is traversed.  In the presence of intralayer correlations, the minimum is either extremely weak or completely absent.  
\end{abstract}

\maketitle

Recent reports of the likely observation of superfluidity with electron-hole pairs in spatially separated electron and hole conducting layers in zero magnetic fields \cite{Burg2018, Wang2019, Ma2021,Gamucci2014, Nguyen2023}, are currently attracting a lot of interest. The spatial separation opens a way to stable superfluids in equilibrium because it suppresses electron-hole recombination \cite{Lozovik1976}.    

Theoretical investigations of these two-layer systems have focused on the electron-hole correlations needed to generate the electron-hole pairs. However, on account of very significant screening effects \cite{Perali2013}, the superfluidity is restricted to low carrier densities, and so correlations between electrons in one layer and correlations between holes in the other layer can be expected to play a significant role.  It is the purpose of this paper to investigate the effect of the correlations acting within each layer on the superfluid properties.

For superfluidity of spatially indirect excitons, the average separation between the excitons is generally much greater than the layer spacing separating the electrons and holes.  The excitons are then well approximated by particles with dipole moments perpendicular to the layers and mutually interacting through repulsive dipole-dipole interactions acting parallel to the layers \cite{Astrakharchik2007, Boning2011}. At the relatively low densities where superfluidity is found \cite{DePalo2002, Wang2019}, kinetic energy effects tend to dominate over the intralayer correlations caused by the dipolar interactions. 
In this case, an expansion of the corrections due to the intralayer correlations will be dominated by the Hartree-Fock contribution. This is in striking contrast to Wigner crystallization in double-layer coulombic systems, where at low densities the intralayer correlations from the Coulomb interactions are dominant over kinetic energy effects \cite{Szymanski1994}.  

In this paper, we investigate the effect of intralayer correlations on superfluidity using the Hartree--Fock approximation. 
The coupled 
mean-field equations for the superfluid gap $\Delta_k$ and layer density $n$ at zero temperature are \cite{Perali2013, Devreese2014, Lumbeeck2020}, 

\begin{eqnarray}
\Delta_k &=& -\frac{1}{S} \sum_{\textbf{k'}}V^{sc}_{eh}(k-k')\frac{\Delta_{k'}}{2E_{k'}}\ ,
\label{gap}\\
n &=& {g_s g_v} \sum_\textbf{k} \frac{1}{2} \left(1-\frac{\varepsilon_{k}-\mu_s}{E_k}\right)\ .
\label{number}
\end{eqnarray}
$E_{k}=\sqrt{\xi_k^2+\Delta_k^2}$ is the excitation energy, $\xi_k=\varepsilon_k-\mu_s$, with $\varepsilon_k$ the single-particle energy band and $\mu_s$ the single-particle chemical potential. $g_s$ and $g_v$ are the spin and valley degeneracies and $S$ is the area of the system.  

\begin{equation}
V^{sc}_{eh}(\mathbf{q})=\frac{V_{eh}(\mathbf{q})-\Pi_a(\mathbf{q})(V_{ee}^2(\mathbf{q})-V_{eh}^2(\mathbf{q}))} 
{1-2(V_{ee}(\mathbf{q})\Pi_n(\mathbf{q})-V_{eh}(\mathbf{q})\Pi_a(\mathbf{q}))
+\mathcal{A}_\mathbf{q}\mathcal{B}_\mathbf{q}} \ ,
\label{Eq:VeffSF} 
\end{equation}
where $V_{ee}(\mathbf{q})=1/q$ is the bare electron-electron (and hole-hole) interaction 
acting within each layer and 
$V_{eh}(\mathbf{q})=-e^{-qd}/q$ the bare electron-hole interaction 
 between layers, where $d$ is the interlayer distance.  
$\Pi_n(\mathbf{q})$ and $\Pi_a(\mathbf{q})$ are the normal and anomalous polarizabilities in the superfluid phase \cite{Lozovik2012,Perali2013}.
For brevity, we write
$\mathcal{A}_\mathbf{q}=V_{ee}^2(\mathbf{q})-V_{eh}^2(\mathbf{q})$ and 
$\mathcal{B}_\mathbf{q}=\Pi^2_n(\mathbf{q})-\Pi_a^2(\mathbf{q})$.

In the Hartree--Fock approximation, the single-particle energy is given by \cite{Giuliani2005, Debnath2017}: 
%
\begin{align}
     \xi^{HF}_\textbf{k}&=\frac{\hbar^2 \textbf{k}^2}{2m}-\mu_s-\Sigma(\textbf{k}) \, ,  \label{Eq:xi}
\end{align}
where
\begin{align}
     \Sigma(\textbf{k})&=\frac{1}{S} \sum_{\textbf{p}}V^{sc}_{ee}(\textbf{p}-\textbf{k})v^2_\textbf{p} \, , \label{Eq:Sigma}
\end{align}
with the Bogoliubov amplitude (density of states)
\begin{equation}
v^2_\textbf{k}=1-u_\textbf{k}^2=\frac{1}{2}\Bigg(1-\frac{\xi^{HF}_\textbf{k}}{E^{HF}_\textbf{k}} \Bigg),     
\end{equation}
\\
The self-consistent static screened electron-electron (hole-hole) interaction within each layer is \cite{Debnath2017},

\begin{align}
V^{sc}_{ee}(\mathbf{q})=\frac{V_{ee}(\mathbf{q})-\Pi_n(\mathbf{q})(V_{ee}^2(\mathbf{q})-V_{eh}^2(\mathbf{q}))} 
{1-2(V_{ee}(\mathbf{q})\Pi_n(\mathbf{q})-V_{eh}(\mathbf{q})\Pi_a(\mathbf{q}))
+\mathcal{A}_\mathbf{q}\mathcal{B}_\mathbf{q}}\label{Eq:VeeeffSF} 
\end{align}

To determine the effect of the Hartree--Fock corrections within the layers, we solve the gap and number equations Eqs.\ \eqref{gap} and \eqref{number} using $\xi^{HF}_\textbf{k}$ for the single-particle energy. The screened interactions, 
Eqs.\ \eqref{Eq:VeffSF} and \eqref{Eq:VeeeffSF}, are modified similarly. 

We take single-particle parabolic bands $\varepsilon_k=\hbar k^2/2m^*$, with equal effective masses $m^*=m_e^*=m_h^*=0.04$.   For the dielectric constant, we use $\epsilon=2$ for double bilayer graphene.  We express lengths in units of the effective Bohr radius, $a_B^*=5.3$ nm, and energies in effective Rydberg $Ry^*= 35$ meV. 
We consider equal electron and hole layer densities, $n=n_e=n_h$, corresponding to an average interparticle spacing in the layers of $r_0=(\pi n)^{-1/2}$.

\begin{figure}[tbp]
\centering
\includegraphics[trim=0.0cm 0.0cm 0.0cm 0.0cm, clip=true, width=\columnwidth]{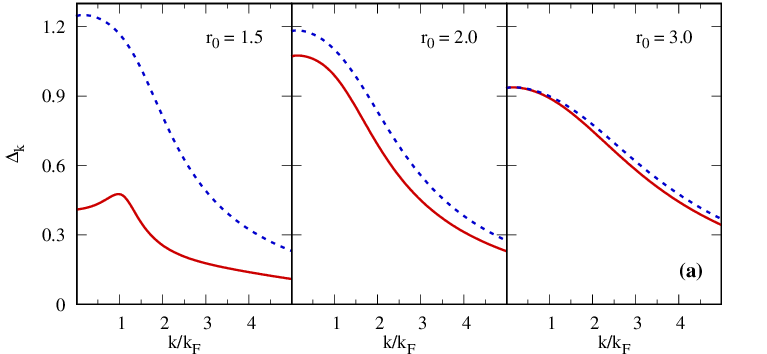}
\includegraphics[trim=0.0cm 0.0cm 0.0cm 0.0cm, clip=true, width=\columnwidth]{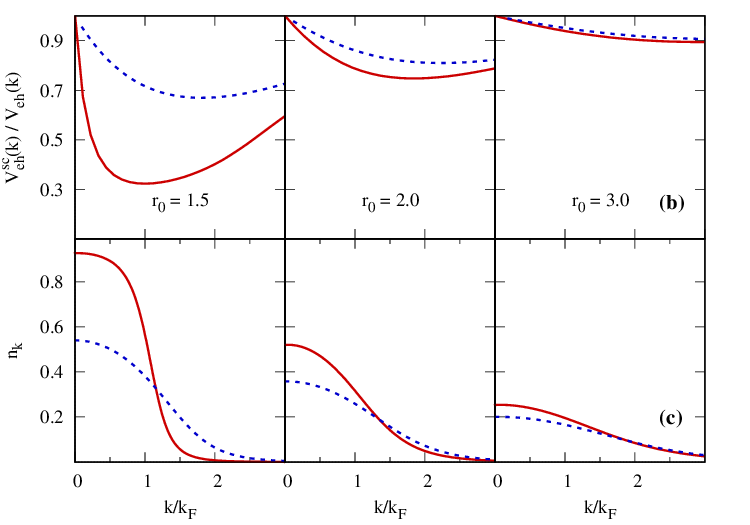}
\caption{(a) Superfluid gap $\Delta_k$ at densities characterized by $r_0$, the average interparticle spacing within each layer. Layer separation $d=0.2$. Solid red: within the mean field including intralayer correlations.  Dashed blue: within the mean-field but neglecting intralayer correlations.
(b) Ratio of self-consistent screened electron-hole attraction $V^{sc}_{eh}(k)$ to the bare attraction $V^{eh}(k)$ for the same densities. 
(c) Corresponding density of states $n_k =v^2_\textbf{k}$. Lengths are in units of the effective Bohr radius and energies are in units of the effective Rydberg (see text).}
\label{Deltak}
\end{figure}

Figure \ref{Deltak}(a) shows the resulting superfluid energy gap $\Delta_k$ for a layer separation $d=0.2$.  The intralayer distance $r_0$ shown spans the full range for superfluidity. Because of strong screening, there is a maximum threshold density for the superfluidity that corresponds to $r_0\simeq 1$.  As the threshold density is approached, we see that the Hartree--Fock correlations have a strong effect on the superfluidity, reducing the gap $\Delta_k$ by as much as a factor of 2. However the effect of the correlations on the superfluidity weakens with decreasing density, and for $r_0\gtrsim 3$, the correlations have negligible effect. 

Figure \ref{Deltak}(b) demonstrates that the suppression of $\Delta_k$ seen at higher densities comes from the effect of the Hartree--Fock correlations weakening the self-consistent electron-hole screened interaction, $V_{eh}^{sc}(\textbf{q})$. This weakening as the density increases is due to Hartree-Fock boosting the number of the low-lying energy states that contribute to the screening (see Fig.\ \ref{Deltak}(c)).

\begin{figure}[t]
\includegraphics[trim=0.0cm 0.0cm 0.0cm 0.0cm, clip=true, width=0.8\columnwidth]{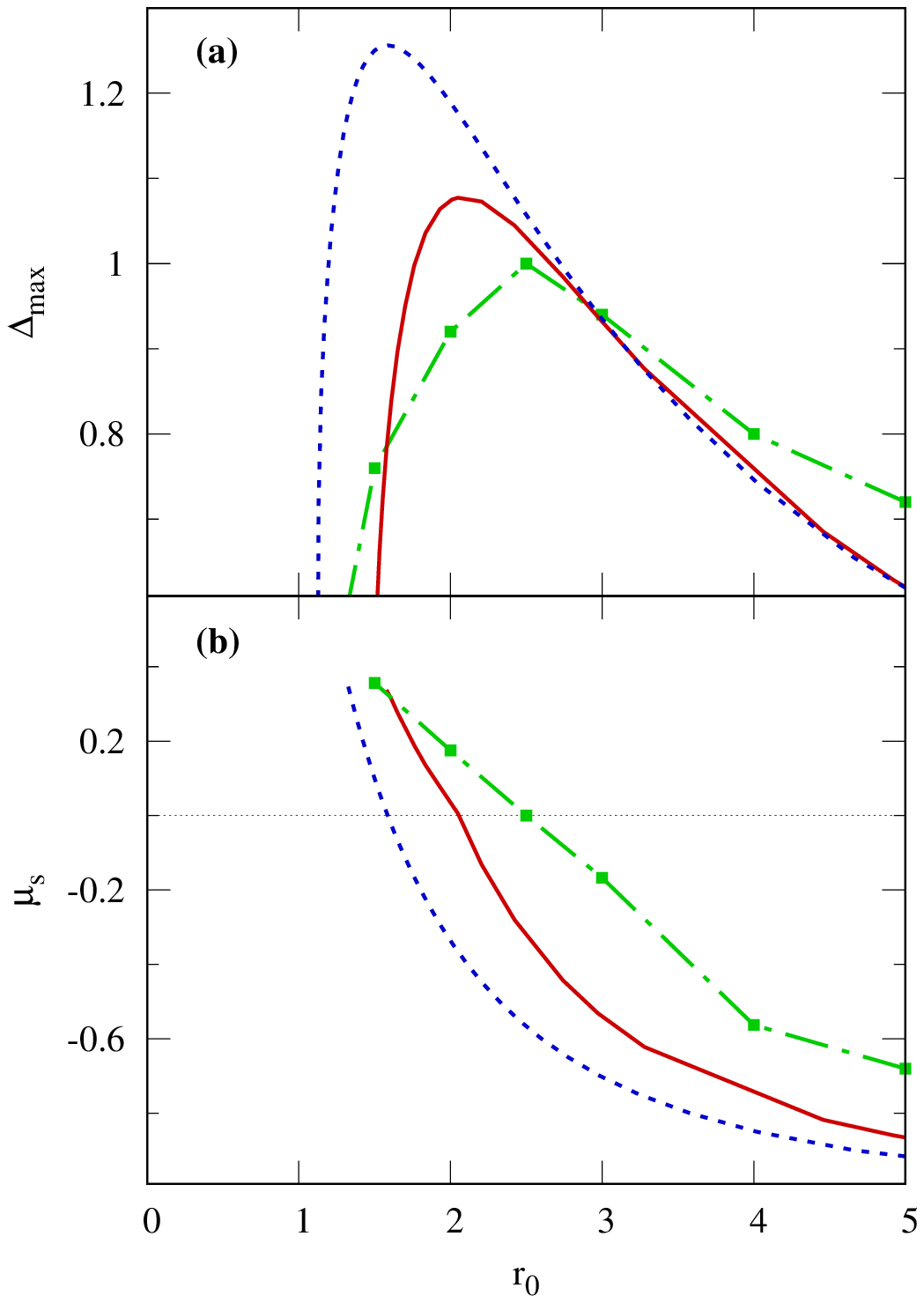}
\caption{(a) Maximum superfluid gap $\Delta_{max}$ as function of  $r_0$, the average interparticle intralayer distance within a layer.  Solid red: within mean field with intralayer correlations included.  Dashed blue: within the mean-field but neglecting intralayer correlations.  Shown for comparison (dash-dot green), the $\Delta_{max}$ from Diffusion Quantum Monte Carlo numerical simulations \cite{LopezRios2018}.  (b) The corresponding single-particle chemical potential $\mu_s$.}
\label{DQMC}
\end{figure}

Figure \ref{DQMC} compares our results with Diffusion Quantum Monte Carlo (DQMC) numerical simulations \cite{LopezRios2018}. We see in Fig.\ \ref{DQMC}(a) that including the Hartree--Fock correlations significantly improves the agreement with DQMC for both the height and position of the maximum of the superfluid peak $\Delta_{max}$. The correlations using static screening push down the threshold density somewhat, but corrections from dynamical screening will act to compensate this \cite{Nilsson2021}. 

Figure \ref{DQMC}(b) compares the single-particle chemical potential $\mu_s$.
We see that the Hartree--Fock corrections are significant and move $\mu_s$ closer to the benchmark DQMC results. 

\begin{figure}[tbp]
\centering
\includegraphics[trim=0.0cm 0.0cm 0.0cm 0.0cm, clip=true, width=\columnwidth]{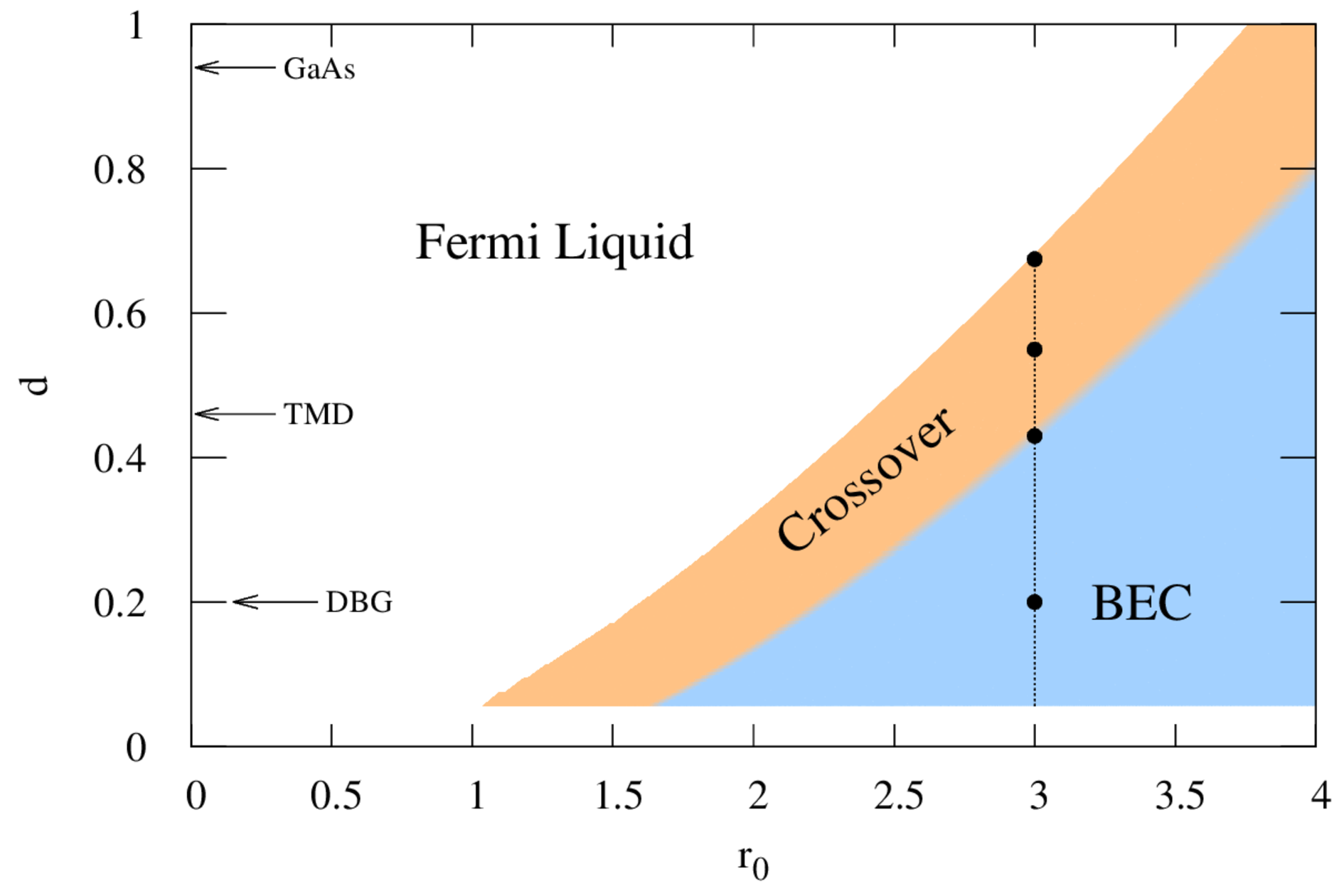}
\caption{Dependence of BEC and BCS-BEC crossover regimes on layer separation $d$ and average interparticle spacing within each layer  $r_0$.  The BCS regime is preempted by strong screening that suppresses superfluidity at small $r_0$ and large $d$.  The smallest separations experimentally attained to date in different material systems are indicated by the arrows on the vertical axis.
Also marked are the points in $r_0$-$d$ phase space used in Fig.\ \ref{Deltadk}. 
}
\label{CF_map}
\end{figure}

An important feature of superfluidity in these electron-hole double-layer systems is that, by tuning the carrier density in the layers $n$ using gate voltages, it is possible experimentally to sweep the superfluidity from a strong-coupled Bose-Einstein condensate (BEC) at the lowest densities, to the intermediate-coupled BCS-BEC crossover regime, through towards the weak-coupled BCS regime \cite{Perali2013,Conti2017}.  
Figure \ref{CF_map} maps out the superfluidity and its regimes at very low temperatures in the $r_0$-$d$ phase space.  We set the boundary between the BEC and the BCS-BEC crossover regimes as the line at which the chemical potential $\mu_s$ changes sign from negative to positive (Fig.\ \ref{DQMC}(b)) \cite{Perali2011,Guidini2014}.   

Indicated for reference on the vertical axis of Fig.\ \ref{CF_map} are the smallest separations experimentally attained to date in Gallium Arsenide (GaAs) double quantum wells  \cite{Chen1987,Eisenstein1990,DasGupta2011}, double layers of bilayer graphene (DBG) \cite{Burg2018}, and double layers of Transition Metal Dichalcogenide (TMD)  \cite{Wang2019,Ma2021}.

\begin{figure}
\centering
\includegraphics[trim=0.0cm 0.0cm 0.0cm 0.0cm, clip=true, width=\columnwidth]{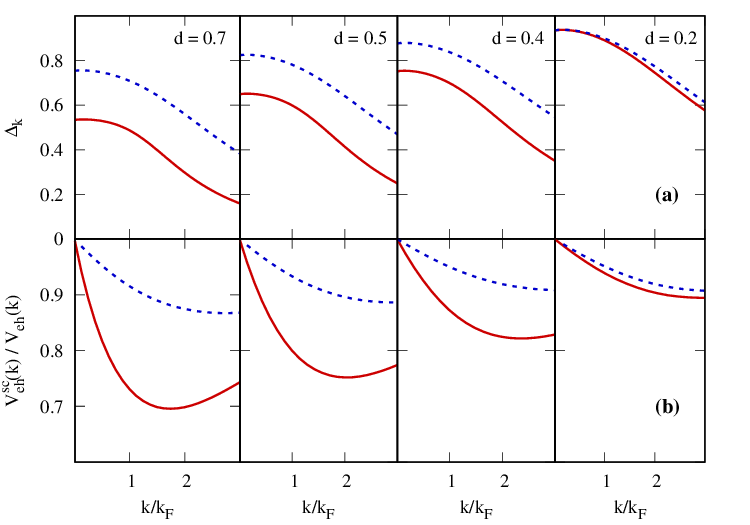}
\caption{(a) Superfluid gap $\Delta(k)$ for a fixed density corresponding to $r_0=3$, at different $d$ points in the BEC and BCS-BEC crossover regimes (refer Fig.\ \ref{CF_map}). Solid red: within mean-field with intralayer correlations included.  Dashed blue: within mean-field but neglecting intralayer correlations. 
(b) Ratio of self-consistent screened electron-hole attraction $V^{sc}_{eh}(k)$ to the bare attraction $V_{eh}(k)$ for the same $r_0$-$d$ points spanning the BEC and BCS-BEC crossover regimes. 
}
\label{Deltadk}
\end{figure}

We compare in Fig.\ \ref{Deltadk}(a) for a fixed value of the layer interparticle spacing $r_0=3$, the evolution of the superfluid gap energy $\Delta_k$ when the Hartree--Fock correlations within the layers are either included or neglected, for different layer separations $d$. The corresponding ($r_0$-$d$) points are marked on the phase diagram, Fig.\ \ref{CF_map}. Figure \ref{Deltadk}(b) compares the corresponding ratios of screened electron-hole attraction $V^{sc}_{eh}(k)$ to the bare attraction $V_{eh}(k)$.

The layer spacing $d=0.2$ lies deep in the BEC regime and Fig.\ \ref{Deltadk}(b) confirms that screening is indeed negligible there.  Since the Hartree--Fock corrections primarily affect the screening, the correlations have almost no effect on $\Delta_k$ for $d=0.2$. However, $d=0.4$ lies on the BCS-BEC crossover boundary, and we see at that point that screening is no longer negligible, and as a consequence, $\Delta_k$ starts to develop a sensitivity to the Hartree--Fock corrections. As $d$ is further increased and the crossover regime is traversed, both the screening and $\Delta_k$ become increasingly sensitive to the Hartree-Fock corrections. By $d=0.7$, the correlations boost the low-lying density of states so much that the screening is strongly enhanced. This in turn strongly suppresses $\Delta_k$.  $d=0.7$ is close to the superfluid threshold where the screening kills the superfluidity.   

\begin{figure}[tbp]
\centering
\includegraphics[trim=0.0cm 0.0cm 0.0cm 0.0cm, clip=true, width=0.9\columnwidth]{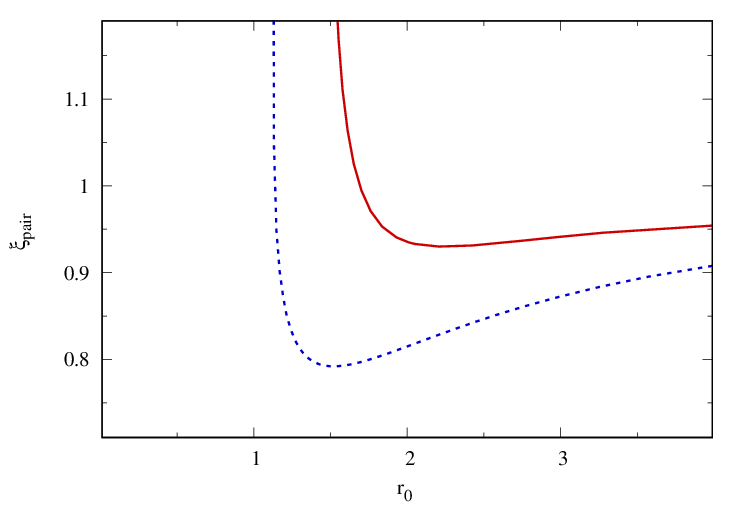}
\caption{The pair size of the exciton $\xi_{pair}$ as a function of $r_0$, the average interparticle spacing in each layer. The layer separation is $d=0.2$.  Solid red: within mean-field with intralayer correlations included.  Dashed blue: within mean-field but neglecting intralayer correlations. 
}
\label{xirs}
\end{figure}

Figure \ref{xirs} compares,  with intralayer correlations included or neglected, the spatial size of the electron-hole pairs \cite{Guidini2014, Pistolesi1994}, 
\begin{equation}
    \xi_{pair}=\Bigg[\frac{\sum_\textbf{k} |\nabla_\textbf{k} u_\textbf{k}v_\textbf{k}|^2}{\sum_\textbf{k} u^2_\textbf{k}v^2_\textbf{k}}\Bigg]^{1/2}\ ,
\end{equation}
as a function of $r_0$ for layer separation $d=0.2$.

Without the intralayer correlations, starting from the low-density BEC regime, $\xi_{pair}$ initially decreases as the density increases. In the BEC regime the pairs act as well-spaced composite bosons interacting primarily through exchange, and so, as the interparticle spacing decreases, exchange effects strengthen causing the pairs to shrink \cite{Andrenacci2000}. In contrast, in the crossover regime the bosonic nature of the pairs is lost because there is significant overlap of the single-fermion wave functions. Thus $\xi_{pair}$ will grow exponentially as the density is further increased. Reference \cite{Andrenacci2000} pointed out that this competing behavior leads to a minimum in $\xi_{pair}$. In Fig.\ \ref{xirs} this minimum is clearly visible when intralayer correlations are omitted.

However,  we see that when intralayer correlations are included, the resulting build up of screening strength with increasing density greatly weakens the shrinkage of $\xi_{pair}$ and effectively eliminates the minimum. Then at higher densities, the very strong screening further weakens the superfluidity, causing the $\xi_{pair}$ to grow exponentially. $\xi_{pair}$ diverges at the threshold density for superfluidity.

\section{Conclusions}
The primary effect of the Hartree--Fock correlations on superfluid properties in the present system is an increase in the strength of screening caused by a boost in the density of the low-lying states.  Screening plays a crucial role in determining superfluid properties because the pairing interaction is long-range \cite{Lozovik2012,Perali2013}, and we find that the strength of the screening can be as much as doubled by the Hartree-Fock corrections. Effects of screening on the superfluidity are negligible in the deep BEC regime \cite{Neilson2014} and therefore Hartree--Fock has minimal effect in that regime, but in the BCS-BEC crossover regime, where screening plays a crucial role in determining the superfluid properties, the increased screening strength results in (i) a diminution of the superfluid gap $\Delta$ by up to a factor of two within the BCS-BEC crossover regime, leading to a better agreement with the DQMC simulations, and (ii) a shift to lower densities of the boundary between the BEC and crossover regimes, which results (iii) in the disappearance of the minimum in the electron-hole pair-size as a function of density.
 \\

{\bf Acknowledgements}\\
We thank S. De Palo and G. Senatore for useful discussions. The work was partially supported by the projects PNRR MUR Project No. PE0000023-NQSTI, G061820N, G060820N, G0H1122N, and by the Flemish Science Foundation (FWO-Vl).

\end{document}